\documentclass[structabstract]{aa}

\usepackage{graphicx}
\usepackage[intlimits]{amsmath}
\usepackage{txfonts}
\usepackage{natbib}
\usepackage{textcomp}
\usepackage{color}
\bibpunct{(}{)}{;}{a}{}{,}

\begin{document}
\title{Modelling accretion in transitional disks}

\author{
	Tobias~W.~A.~M\"uller\inst{\ref{inst1}}
	\and Wilhelm~Kley\inst{\ref{inst1}}
}

\institute{
	Institut f\"ur Astronomie \& Astrophysik, 
	Universit\"at T\"ubingen,
	Auf der Morgenstelle 10,
	72076 T\"ubingen,
	Germany
	\label{inst1} \\
	\email{Tobias\_Mueller@twam.info}
}

\date{Received 2013-08-19 / Accepted 2013-10-14	}

\abstract
{Transitional disks are protoplanetary disks around young stars that display inner holes in the dust distribution
within a few au, that are accompanied by some gas accretion onto the central star. 
These cavities could possibly be created by the presence of one or more massive planets that
opened a large gap or even cleared the whole inner region.}
{ If the gap is created by planets and gas is still present in it, then there should be
a flow of gas past the planet into the inner region.
It is our goal to study in detail the mass accretion rate into this planet-created gap in transitional disks and
in particular the dependency on the planet's mass and the thermodynamic properties of the disk.}
{We performed 2D hydrodynamical simulations using the grid-based \texttt{FARGO} code for disks
with embedded planets.
We added radiative cooling from the disk surfaces, radiative diffusion in the disk midplane,
and stellar irradiation to the energy equation to have more realistic models.}
{The mass flow rate into the gap region depends, for given disk thermodynamics, non-monotonically
on the mass of the planet. Generally, more massive planets open wider and deeper gaps
which would tend to reduce the mass accretion into the inner cavity. However, for larger mass planets
the outer disk becomes eccentric and the mass flow rate is enhanced over the low mass cases.
As a result, for the isothermal disks the mass flow is always comparable to the expected mass flow of
unperturbed disks $\dot{M}_\mathrm{d}$, while for more realistic radiative disks the mass flow is very small for
low mass planets ($\leq 4\,M_\mathrm{jup}$) and about 50\,\% of $\dot{M}_\mathrm{d}$ for larger planet masses.
The critical planet mass that allows the disk to become eccentric is much larger for radiative disks than for purely isothermal cases.
}
{Massive embedded planets can reduce the mass flow across the gap considerably, to values
of about an order of magnitude smaller than the standard disk accretion rate, and can be responsible for opening large
cavities. The remaining mass flow into the central cavity is in good agreement with the observations. }

\maketitle

\section{Introduction}
\label{sec:introduction}

The early stages of planet formation take place in protoplanetary disks around young stars and the
growing and evolving planets shape the structure of the disks. One important observational goal is the detection
of signatures that give direct hints of the presence of planets in such disks.
One class of systems that have been linked to planets are the so-called transitional disks that show 
a deficiency in the IR excess in their spectral energy distribution (SED) and/or show an extended inner hole in the disk in
direct images.  
In recent years many of these transitional disks have been discovered, for example
by the Spitzer Space Telescope \citep{2010ApJ...708.1107M, 2013ApJ...769..149K}. 
The occurrence rate of transitional disks rises monotonically with age from about 1\,\% at an age of one million years to
about 20\,\% at 8 million years \citep{2010ApJ...708.1107M} and so they are potentially very interesting indicators
of the late phase in the planet formation process.
Disks with gaps or completely cleared inner cavities usually fshow mass accretion onto the star,
but with accretion rates $\dot{M}$ typically an order of magnitude smaller than that of a continuous disk \citep{2013ApJ...769..149K}.
As there is a clear decrease in $\dot{M}$ from continuous disks over disks with gaps to disks with cavities,
\citet{2013ApJ...769..149K} suggest that the presence of sub-stellar companions (planets) may be the most likely explanation of 
the properties of transitional disks.
 
Growing planets can open gaps within the disk after accumulating a sufficient amount of mass, and so they constitute
a natural possible explanation \citep{1993prpl.conf..749L}. Indeed, single planet simulations of embedded planets could
show the formation of large inner cavities with a reduced mass flow through the inner hole 
\citep{2004ApJ...612L.137Q,2006ApJ...640.1110V}. In these simulations, a massive Jupiter-like planet
opened a gap in the disk and the inner cavity was cleared out rapidly through mass flow through the inner boundary onto the
star.  While it was possible to produce a deep cavity, it has been pointed out that this may be due to the special
outflow boundary condition used at the inner edge \citep{2007A&A...461.1173C}. 
Additional clearing of the inner disk in the case of one embedded planet region may also be mediated
by disk photo-evaporation \citep{2009ApJ...704..989A}, which has been investigated in simulations by \citet{2013MNRAS.430.1392R}.
Here, the timescale of the evaporation process has to be
chosen such that it is compatible with the observed long life times of accretion in T Tau disks \citep{2011ApJ...729...47Z}.
Because it is known that multiple planets can clear out much deeper and wider gaps than single
planets \citep{2000MNRAS.313L..47K}, it has been suggested that the large gaps and inner cavities
may be caused by the presence of multiple (up to four) planets \citep{2011ApJ...738..131D,2011ApJ...729...47Z}.
In this scenario, however, mass transfer has to occur across the planets and the system has to be dynamically
stable over long timescales.

The recent observations presented by \citet{2012ApJ...760L..26M} that show a kind of spiral arm connecting the outer
disk through the cleared area with the star strengthened these hints of planets within transitional disks.
In either case, multiple planets or a single planet, the important question is, how strong is the mass flow from the outer
disk across the planet into the inner cavity onto the star?	

Even though the option of multiple planets has been suggested as a possible cause for the observed properties of 
transitional disks, we will nevertheless focus in this paper of the effect of single embedded planets in the
disk. As pointed out above, even in the presence of multiple planets it is necessary for mass to flow
from the outer disk into the inner gap region to provide for the observed accretion signatures onto the star.
Here, we focus on the dynamics of single massive planets that are surrounded by an outer accretion disk and no inner
counterpart. For smaller mass planets ($< M_\mathrm{jup}$) the mass accretion rate across a gap has been analysed
by \citet{2006ApJ...641..526L}. This single planet could be considered the outermost in a multi-planet
system. Our simulations will give important information about the mass flow from the outer disk into
this planetary system.

We are particularly interested in the amount of mass that can flow across massive planets into the inner
cavity. For this purpose we have performed a detailed analysis of this process considering planets of different masses.
Additionally, we have improved on the thermodynamics of the disk and consider isothermal as well 
as radiative disks with and without irradiation from the central star.

In Sect.~\ref{sec:model} we describe our physical and numerical modelling of the process. The standard model is
described in Sect.~\ref{sec:standard}. In Sect.~\ref{sec:isothermal} we present the results of the isothermal
runs for different planet masses together with some numerical tests. This is followed in Sect.~\ref{sec:radiative}
by the radiative results that are compared to even more extended models that include irradiation in Sect.~\ref{sec:comparison}.
 
\section{Model set-up}
\label{sec:model}

\allowdisplaybreaks

\subsection{Physics and equations}

We assumed an infinitesimally thin disk around the star and therefore solved the vertically integrated versions of the
hydrodynamical equations. For the coordinate system we chose cylindrical coordinates $(r, \varphi, z)$,
centred on the star where the disk lies in the equatorial $ z = 0$ plane. 

The vertically integrated versions of the continuity equation and the equations of motions in the $r$--$\varphi$ plane can be
found in \citet[Section 2.1]{2012A&A...539A..18M}. Here, we included an energy equation to allow for a 
more realistic thermodynamic treatment of the disk.
The vertically integrated energy equation reads
\begin{align}
	\label{eq:energy_equation}
	\frac{\partial e}{\partial t} + \nabla \cdot \left(e \vec{v} \right) &= -p \nabla \cdot \vec{v} + Q_{+} - Q_{-} \ ,
\end{align}
where $ e $ is the internal energy density, $ Q_{+} $ the heating source term, and $ Q_{-} $ the cooling source term.

The cooling term $ Q_{-} $ is given by the vertically integrated divergence of the radiative flux,
\begin{align}
	F &= - \frac{4 \sigma_\mathrm{R}}{3 \rho \kappa} \nabla T^4 \ ,
\end{align}
where $ \sigma_\mathrm{R} $ is the Stefan-Boltzmann constant, $ \rho $ the density, $ \kappa $ the Rosseland mean opacity 
and $ T $ the temperature. It can be written as
\begin{align}
	Q_{-}	=& - \int_{-\infty}^\infty \nabla \cdot F\,dz = - \int_{-\infty}^\infty \nabla \cdot \left( -\frac{16 \sigma_\mathrm{R}}{3 \rho \kappa} T^3 \nabla T \right) \, dz \nonumber \\
	=& - \int_{-\infty}^\infty \frac{\partial}{\partial z}  \left(-\frac{16 \sigma_\mathrm{R}}{3 \rho \kappa} T^3 \frac{\partial}{\partial z} T \right) \, dz \nonumber \\
	 &- \int_{-\infty}^\infty \left[ \frac{1}{r} \frac{\partial}{\partial r}  r \left(-\frac{16 \sigma_\mathrm{R}}{3 \rho \kappa} T^3 \frac{\partial}{\partial r} T \right) + \frac{1}{r} \frac{\partial}{\partial \varphi} \frac{1}{r} \left(-\frac{16 \sigma_\mathrm{R}}{3 \rho \kappa} T^3 \frac{\partial}{\partial \varphi} T \right)  \right] \, dz \nonumber \\
	\approx & - Q_{\mathrm{rad}} - 2 H \, \nabla \cdot \left( K \nabla T \right) \label{eq:q-} \ .
\end{align}
Here $ Q_\mathrm{rad} $ describes the radiative losses from the disk surfaces which is calculated using suitably averaged 
opacities \citep[Sect. 2.1]{2012A&A...539A..18M},
and the second term corresponds to the radiative diffusion in plane of the disk; 
$ H = c_\mathrm{s}/\Omega_\mathrm{K}$ is the pressure scale height
given by the midplane sound speed, $c_\mathrm{s}$, and the Keplerian rotational angular velocity, $\Omega_\mathrm{K}$, of the disk.
The coefficient of the radiative diffusion in the $r$--$\varphi$ plane is given by
$ K = - \frac{16 \sigma_\mathrm{R}}{3 \rho \kappa} T^3 $.

The heating term $ Q_{+} $ consists of the viscous dissipation $ Q_\mathrm{visc}$
and the irradiation from the star $ Q_\mathrm{irrad} $. The stellar irradiation to one surface of the disk can be approximated by
\begin{align}
	Q_\mathrm{irrad} = \beta \sigma_\mathrm{R} T_\mathrm{star}^4 \left(\frac{R_\mathrm{star}}{r}\right)^2
\end{align}
\citep{2004A&A...423..559G}, where $ T_\mathrm{star} $ is the effective photospheric temperature of the star
and $ R_\mathrm{star} $ the radius of the star.
The factor $ \beta $ accounts for the non-perpendicular impact of the irradiation from the star onto the disk.

\subsection{Numerical considerations}

We used the adiabatic version of the \texttt{FARGO} code \citep{2000A&AS..141..165M, baruteauthesis}. For the radiative cooling $ Q_\mathrm{rad} $ from Eq. \ref{eq:q-} we used our implementation from \citet{2012A&A...539A..18M}.
The radiative diffusion source term is solved separately after the rest of the source-terms as a flux-limited diffusion
equation for the temperature \citep{1989A&A...208...98K,2008A&A...487L...9K}.
We solve it implicitly using a successive over-relaxation (SOR) method.
For the stellar irradiation we implemented the approximation by \citet{2004A&A...423..559G}.

The planet in the disk has a Plummer type gravitational potential to account for the vertical extent of the disk and to avoid numerical problems of a point-mass potential. We used a smoothing value of $ \epsilon = 0.6\,H $ as
this describes the vertically averaged forces very well \citep{2012A&A...541A.123M}.

To avoid numerical problems the density cannot fall below a minimum value of $ \Sigma_\mathrm{floor} = 10^{-7} \cdot \Sigma_0 $,
where $\Sigma_0$ is the reference density at $r = r_\mathrm{jup} = 5.2$\,au, and the temperature is always at least 
$ T_\mathrm{floor} = 3\,\mathrm{K} $, which is about the temperature of the cosmic background radiation.

Because the disk has an abrupt end at the inner edge of the computational domain, the stellar irradiation produces here 
artificial high temperatures because of the large amount of energy deposited.
To account for this we introduced an estimated horizontal optical depth $ \tilde{\tau} $ by rescaling the vertical optical depth
$ \tau $ \citep[see][Eq.~12]{2012A&A...539A..18M} by the radial extent of the corresponding grid cell
\begin{align}
	\tilde{\tau}_{i,j} = \tau_{i,j} \frac{r_{i+1}-r_{i}}{\left(h r\right)_{i,j}} \,,
\end{align}
where $ h = H/r $ is the aspect ratio of the disk. If $ \tilde{\tau}_{i,j} < 1 $ we set $ Q_\mathrm{irrad} = 0 $.
This simple procedure keeps the temperatures within the gap region regular and allows for standard heating in the outer
parts of the disk.

\subsection{Boundary conditions}

To maintain a given disk structure in the outer parts of the disk 
we implemented the damping mechanism by \citet{2006MNRAS.370..529D},
where specified quantities like radial velocity and angular velocity are damped towards their initial values. 
This damping is described by
\begin{align}
	\frac{d\xi}{dt} = - \frac{\xi - \xi_0}{\tau} R(r)^2 \ ,
\end{align}
where $ \xi \in \{ \Sigma, v_r, v_\varphi, e \} $, $ \tau $ is the damping timescale, and $ R(r) $ is a linear ramp-function 
rising from $ 0 $ to $1$ from the damping star radius to
the outer radius of the computational domain. To be able to measure the mass loss rate in our models 
we only damped the velocity components within $ 0.9\,r_\mathrm{max} $ to $ r_\mathrm{max} $ using a time factor of
$ \tau = 3 \times 10^{-2} \cdot 2 \pi \, \Omega_\mathrm{K}(r_\mathrm{max})^{-1} $.

In addition to the damping we used a reflecting boundary condition at the outer edge so that no mass could escape through
the outer boundary. For the inner edge we used a zero-gradient outflow boundary condition and measured the amount of mass
lost through this boundary to calculate the mass accretion rate onto the star.
The details of the implementation of the boundary conditions can be found in \citet[Section 2.2]{2012A&A...539A..18M}.

\section{The standard model}
\label{sec:standard}

Our model consists of a planet with a varying mass of $ 1 $ to $ 16 $ Jupiter masses ($M_\mathrm{jup}$) orbiting a Sun-like star
with a semi-major axis of $ 1\,r_\mathrm{jup} = 5.2\,\mathrm{au} $. 
The  planet is fixed in its circular orbit and therefore cannot change its orbital parameters during the simulation.
Beyond $r_\mathrm{jup}$ the star is surrounded by a gaseous disk that extends up to $ r_\mathrm{max} = 5\,r_\mathrm{jup} $, the outer edge
of the computational domain. The inner radius depends on the model and ranges from $ 0.7 $ to $ 0.9\,r_\mathrm{jup} $, where
we chose for the main part of our models $ r_\mathrm{min} = 0.7$.
The gas disk is set up initially as if there were no planet with a simple surface density power-law profile of $ r^{-0.5} $
 and a temperature power-law profile of $ r^{-1} $. To prevent strong shock waves in the initial phase of the simulations
the planet's mass is ramped up slowly over $50$ orbits. All the models were run for a total simulation time of $ 5000 $ planetary orbits.
For the viscosity we used a constant kinematic viscosity $ \nu = 10^{15} \mathrm{cm}^2\,\mathrm{s}^{-1} $ that corresponds to an $\alpha$-value
of about $ 4 \times 10^{-3} $ at the reference radius.
Table \ref{tab:parameters} summarizes the parameters of the standard model.

\begin{table}
	\caption{
		\label{tab:parameters}
		Parameters of the standard disk model.
	}
	\centering
	\renewcommand\arraystretch{1.2}
	\begin{tabular}{ll}
		\hline 
		Star mass ($ M_\mathrm{star} $)	& $ 1\,M_{\sun} $ \\
		Star radius ($ R_\mathrm{star} $)	& $ 1\,R_{\sun} $ \\
		\hline 
		Planet mass ($ M_\mathrm{p} $)		& $ 4\,M_\mathrm{jup} $ \\
		\hline
		Adiabatic index ($ \gamma $)		& $ \frac{7}{5} $ \\
		Mean-molecular weight ($ \mu $)	& $ 2.35 $ \\
		\hline
		Surface density ($\Sigma_0 $)		& $ 888.723\,\mathrm{g\,cm^{-2}} $ \\
		Viscosity ($\nu$)				& $1 \times 10^{15}\,\mathrm{cm^2\,s^{-1}}$ \\
		\hline
		Initial density profile ($ \Sigma $)	& $ \propto r^{-1/2} $ \\
		Initial temperature profile ($ T $)	& $ \propto r^{-1} $ \\
		Initial disk aspect ratio ($ H/r $)	& $ 0.05 $ \\
		\hline
		Grid ($N_r \times N_\varphi$)	& $ 256 \times 1024 $ \\
		Computational domain ($ r_\mathrm{min} $ -- $ r_\mathrm{max} $) & $0.7$ -- $5\,r_\mathrm{jup} $ \\
		\hline
	\end{tabular}
\end{table}

\section{Isothermal simulations}
\label{sec:isothermal}

For the first set of simulations we used a locally isothermal approach keeping the initial temperature stratification throughout the
whole simulation fixed.  We assumed a constant aspect ratio $ H/r =  0.05 $, that corresponds to a temperature 
profile of $ r^{-1} $ with $ 137\,\mathrm{K} $ at the inner edge and $24\,\mathrm{K}$ at the outer edge of the disk.
Similar simulations have been performed by \citet{2006A&A...447..369K} for a constant surface
density profile; they damped the surface density at the outer edge of the disk.

\subsection{Inner radius dependency}

\begin{figure}
	\centering
	\includegraphics[width=\columnwidth]{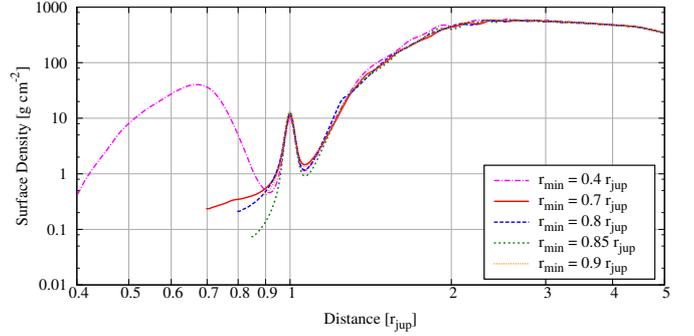} 
	\caption{
		\label{fig:sigma_radius_isothermal_rmin_5000_log}
		Azimuthally averaged radial profile of surface density of the isothermal set-up after $ 5000 $ orbits with a planet of $M_\mathrm{p} = 4\,M_\mathrm{jup} $.
	}
\end{figure}

\begin{figure}
	\centering
	\includegraphics[width=\columnwidth]{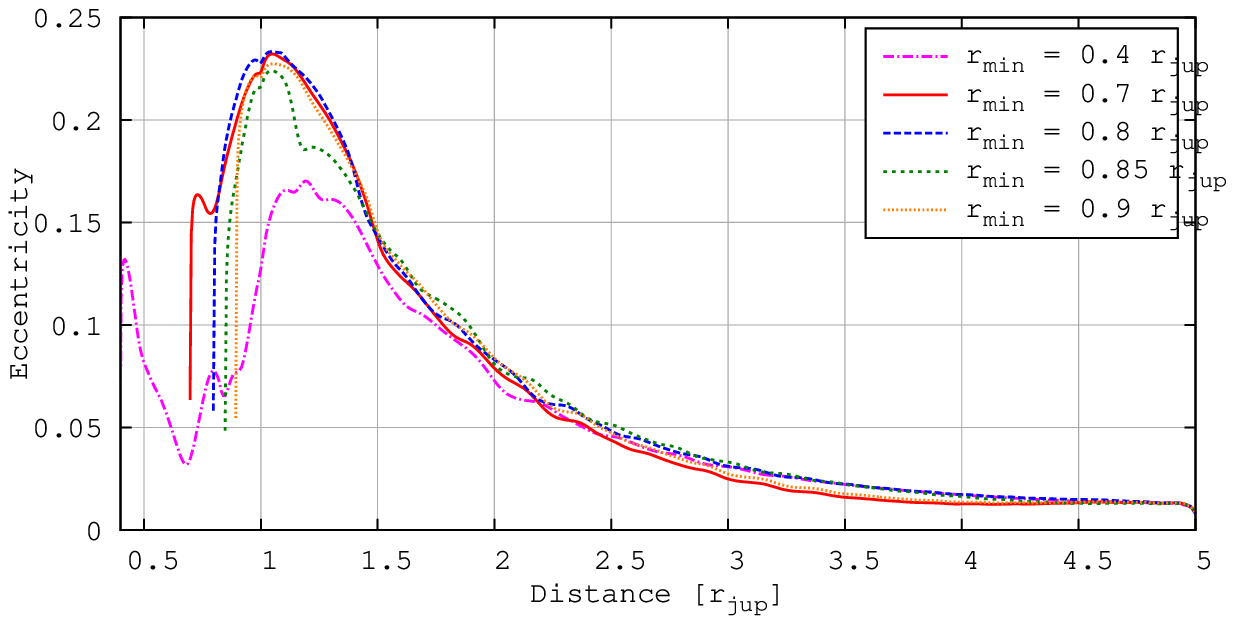} 
	\caption{
		\label{fig:eccentricity_radius_isothermal_rmin_5000}
		Azimuthally averaged radial profile of eccentricity of the isothermal set-up after $ 5000 $ orbits with a planet of $ M_\mathrm{p} = 4\,M_\mathrm{jup} $.
	}
\end{figure}

\begin{figure}
	\centering
	\includegraphics[width=\columnwidth]{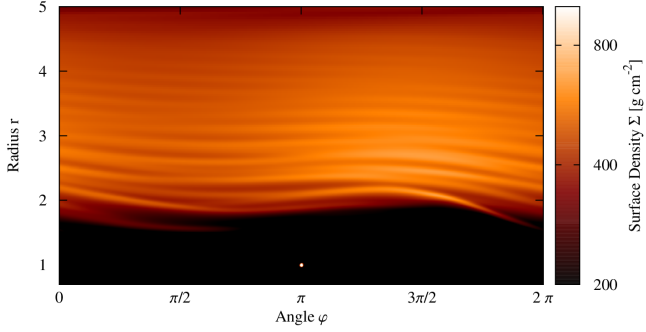} 
	\caption{
		\label{fig:sigma_isothermal_5000}
		Surface density of the isothermal set-up after $ 5000 $ orbits
		with a planet of $ M_\mathrm{p} = 4\,M_\mathrm{jup} $ and an inner radius of $ r_\mathrm{min} = 0.7\,\mathrm{jup} $. 
		The planet is located at the angle of $ \pi $ in this $r-\phi$ representation.
	}
\end{figure}

In this first set of simulations, a fixed planet mass of $ M_\mathrm{p} = 4 M_\mathrm{jup}$ was assumed,
while the location of the inner radius was varied to check its influence.
The $ L_1 $ point lies at about $ 0.896 $\,au and so all radii were chosen in such a way that the Roche-Lobe of the planet is fully included in the computational domain.
\citet{2006A&A...447..369K} showed, that for planets with $ M_\mathrm{p} \geq 3 M_\mathrm{jup} $ the disks become eccentric after a very long time ($> 1000$ orbits).
Figures~\ref{fig:sigma_radius_isothermal_rmin_5000_log} and  \ref{fig:eccentricity_radius_isothermal_rmin_5000} show the azimuthally 
averaged surface density and disk eccentricity profiles after $ 5000 $ orbits for $ M_\mathrm{p} = 4 M_\mathrm{jup}$ and five different inner
radii. For the eccentricity we calculated first a mass weighted average of each grid cell and then averaged over the azimuthal direction.
We note that the models are in different phases of their eccentricity oscillation as shown in Fig.~\ref{fig:eccentricity_time_isothermal_rmin}.
The gap opened by the planet at $ r = 1\,r_\mathrm{jup} $ is clearly visible and the gas in the vicinity has eccentric orbits with an eccentricity of up to $ 0.22 $. 
The eccentricities for all models are very similar. The model with the smallest inner
radius shows a slightly smaller eccentricity just outside of the planet inside the gap.
This does not influence the mass flow across the planet, however.
The eccentricity can also be noticed in the surface density plot show in Fig.~\ref{fig:sigma_isothermal_5000}. In this $ r$-$\varphi$ plot, circular motions are represented by horizontal lines.
However, the gas at the outer edge of the gap has wavelike perturbations which
visualize the eccentricity of its orbit. This is similar to Fig.~1 of \citet{2006A&A...447..369K} where the outer disk also becomes eccentric.

\begin{figure}
	\centering
	\includegraphics[width=\columnwidth]{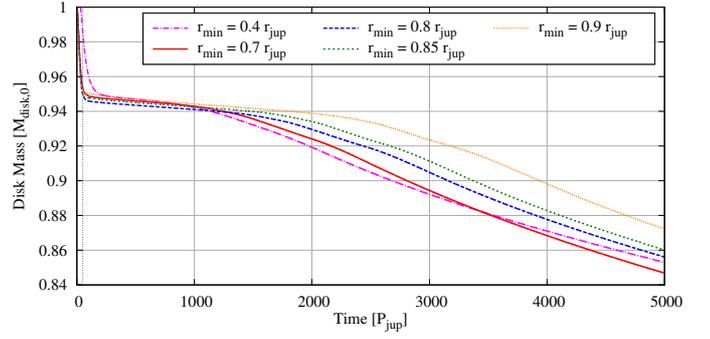} 
	\caption{
		\label{fig:mass_time_isothermal_rmin}
		Total disk mass over time for the isothermal set-up with a planet mass of $ M_\mathrm{p} = 4\,M_\mathrm{jup} $.
		The vertical dotted line at $ t = 50\,P_\mathrm{jup}$ marks the time when the planet has reached its full mass.
	}
\end{figure}

\begin{figure}
	\centering
	\includegraphics[width=\columnwidth]{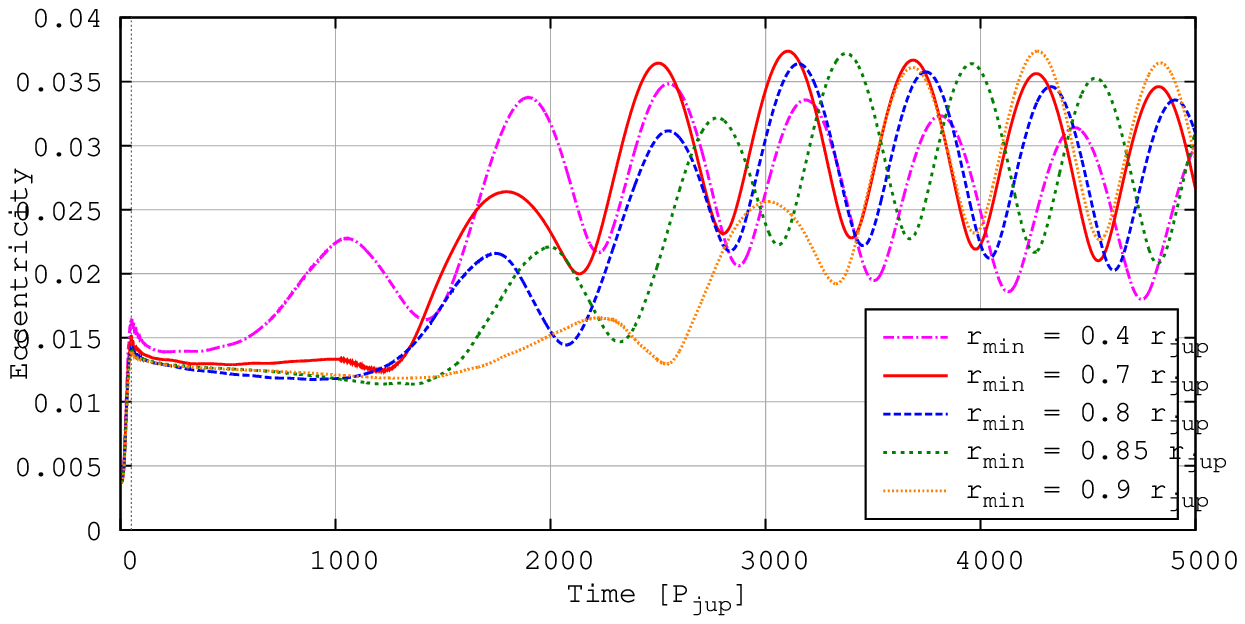} 
	\caption{
		\label{fig:eccentricity_time_isothermal_rmin}
        Global disk eccentricity over time for the isothermal set-up with a planet mass of $ M_\mathrm{p} = 4\,M_\mathrm{jup} $.
		The vertical dotted line at $ t = 50\,P_\mathrm{jup}$  marks the time when the planet has reached its full mass.
	}
\end{figure}

These quasi-stationary states of the disks need a rather long time to establish. 
Figure~\ref{fig:mass_time_isothermal_rmin} shows the disk mass evolution and Fig.~\ref{fig:eccentricity_time_isothermal_rmin} the
disk eccentricity evolution of the disks over the total simulation time of $ 5000 $ orbits.
The planet's mass is ramped up during the first $ 50 $ orbits which is indicated by the dotted vertical line. 
For about the first $1000$ orbits (depending on the inner radius), the disk evolution is very smooth with a nearly constant mass content and small eccentricity.
After this time the disk becomes eccentric and thereby loses more mass through the inner boundary. 
The simulations with smaller computational domain (larger $r_\mathrm{min}$) need some more time
to develop the disk eccentricity, but after that display a similar evolution.

\begin{figure}
	\centering
	\includegraphics[width=\columnwidth]{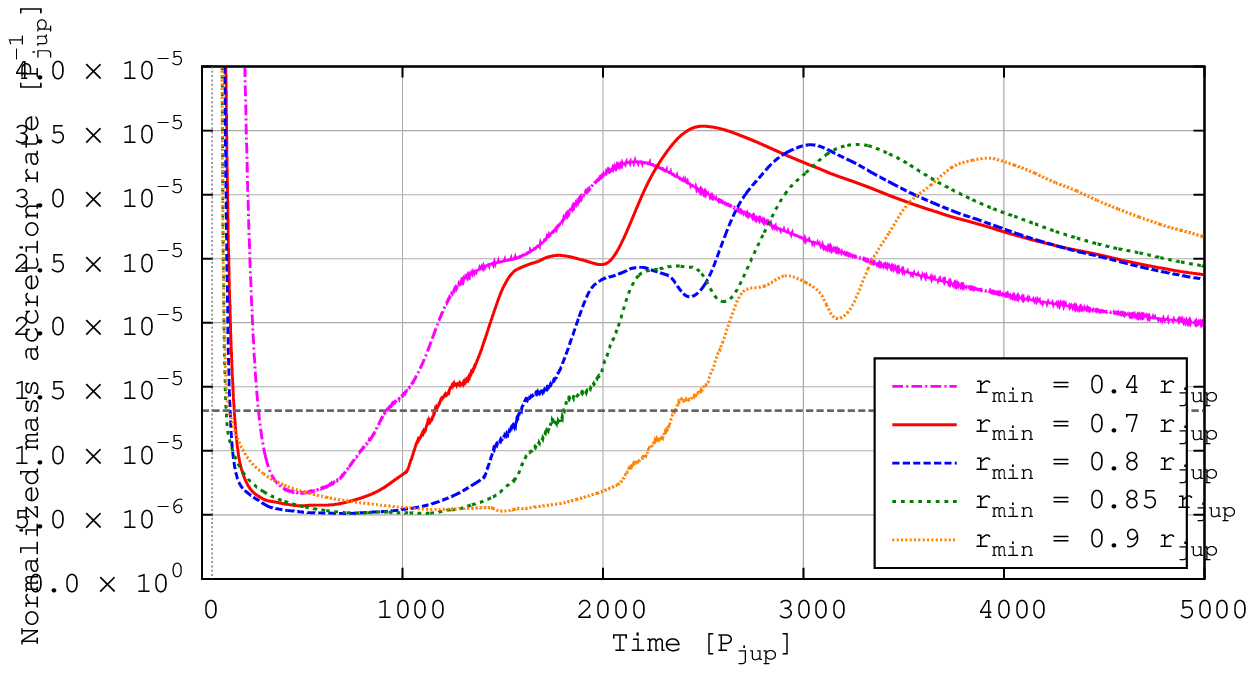} 
	\caption{
		\label{fig:massdot_time_isothermal_rmin}
	Normalized mass accretion rate over time for the isothermal set-up with a planet mass of $ M_\mathrm{p} = 4\,M_\mathrm{jup} $.
		The values are plotted as a moving average over $ 50 $ orbits to remove jitter.
		The vertical dotted line at $ t = 50\,P_\mathrm{jup}$ marks the time when the planet has reached its full mass and the horizontal dashed line is the gas accretion rate given by Eq. (\ref{eq:accretionrate}).
	}
\end{figure}

Figure~\ref{fig:massdot_time_isothermal_rmin} shows the normalized mass accretion rate
through the inner boundary of the computational domain using a moving average over $ 50 $ orbits.
The normalized mass flow $-\dot{M}(t)/M_\mathrm{disk}(t)$ is shown where the actual mass of the disk is used, as the surface density $ \Sigma $ and therefore the disk mass
 $ M_\mathrm{disk} $ cancel out of the equations in locally isothermal simulations.
During the first $\sim 1000 $ orbits the mass accretion rate is almost constant at about $ 5 \times 10^{-6}\,M_\mathrm{disk}\,P^{-1}_\mathrm{jup} $.
However, as soon as the disk becomes eccentric (see Fig.~\ref{fig:eccentricity_time_isothermal_rmin}) the mass accretion
rate starts to increase as more mass is crossing the planetary orbit and flows through the inner boundary at $r_\mathrm{min}$.

The initial mass loss through the inner boundary is smaller
than the standard mass accretion rate of stationary, viscous accretion disks as given by 
\begin{align}
	\label{eq:accretionrate}
	\dot{M}_\mathrm{disk} &= 3 \pi \nu \Sigma_{0} = 8.37 \times 10^{18} \mathrm{g\,s^{-1}} = 1.33 \times 10^{-7} M_{\sun} \mathrm{a}^{-1} \nonumber \\&= 1.58 \times 10^{-6}  M_{\sun} P^{-1}_\mathrm{jup} \,.
\end{align}
After the disk becomes eccentric, it seems to converge against the value, which is indicated by the horizontal dashed line.

We note that all the figures with normalized mass accretion rates are displayed in units of $ P^{-1}_{\mathrm{jup}} $, whereas accretion rates are usually given in $ M_{\sun}\,\mathrm{a}^{-1} $.
The $ 0.7\,r_\mathrm{jup} $ models have an initial disk mass of $ 0.12\,M_{\sun} $, and because the disks lose less than 20\,\% of their mass during the simulation time of $ 5000$ orbits (see Fig.~\ref{fig:mass_time_isothermal_rmin}),
one can get a rough estimate of the mass accretion rate in $ M_{\sun}\,\mathrm{a}^{-1} $ by dividing the values in the figures by a factor of
\begin{align}
	\label{eq:factor}
	\frac{1}{M_\mathrm{disk}} &= \frac{1}{0.12\,M_{\sun}} \frac{11.86\,\mathrm{a}}{P_\mathrm{jup}} = 98.86\,\mathrm{a}\,M_{\sun}^{-1}\,P_\mathrm{jup}^{-1} \approx 100 \,.
\end{align}

The shapes of the time-dependent curves (Figs.~\ref{fig:mass_time_isothermal_rmin} -- \ref{fig:massdot_time_isothermal_rmin}) appear to be shifted in time, because
the disks with the smaller inner computational radii need more time to loose the mass in the outer disk. This leads to a surface density enhancement at
the outer edge of the gap and subsequently to stronger eccentricity excitations. 

\subsection{Planet mass dependency}

\begin{figure}
	\centering
	\includegraphics[width=\columnwidth]{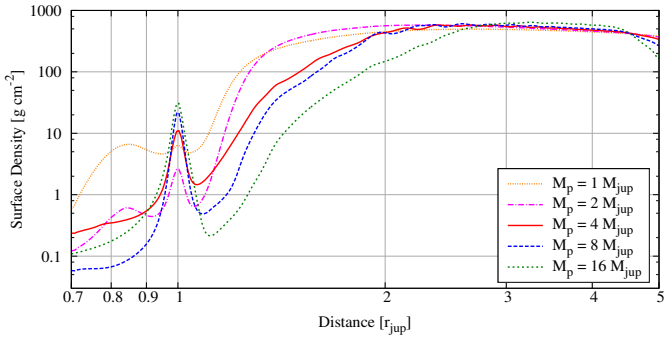} 
	\caption{
		\label{fig:sigma_radius_isothermal_mplanet_5000_log}
		Azimuthally averaged radial profile of surface density of the isothermal set-up after $ 5000 $ orbits with an inner radius of the computational domain of $ r_\mathrm{min} = 0.7\,r_\mathrm{jup} $.
	}
\end{figure}

\begin{figure}
	\centering
	\includegraphics[width=\columnwidth]{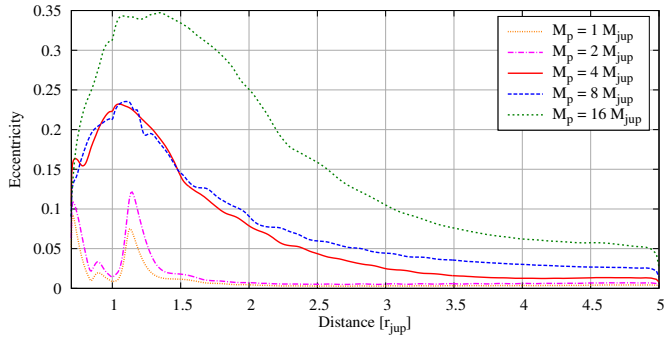} 
	\caption{
		\label{fig:eccentricity_radius_isothermal_mplanet_5000}
		Azimuthally averaged radial profile of eccentricity of the isothermal set-up after $ 5000 $ orbits with an inner radius of the computational domain of $ r_\mathrm{min} = 0.7\,r_\mathrm{jup} $.
	}
\end{figure}

The final state of the simulations are not strongly dependent on the minimum radius of the computational domain.
So we chose $ r_\mathrm{min} = 0.7\,r_\mathrm{jup} $ for the following simulations to compare different planet masses.
As shown by \citet{2006A&A...447..369K}, the outer disk turns eccentric for planetary masses larger than about $ 3\,M_\mathrm{jup} $.
So we varied the planet mass from $ 1\,M_\mathrm{jup} $ to $ 16\,M_\mathrm{jup} $ by doubling it each step.
Figure~\ref{fig:sigma_radius_isothermal_mplanet_5000_log} and \ref{fig:eccentricity_radius_isothermal_mplanet_5000} show again
the radial profiles of the surface density and the eccentricity at the end of simulation after $ 5000 $ orbits.
We note that the models are in different phases of their global eccentricity oscillation as shown in Fig.~\ref{fig:eccentricity_time_isothermal_mplanet}. For example,
in Fig.~\ref{fig:eccentricity_radius_isothermal_mplanet_5000} the $ 16\,M_\mathrm{jup} $ model shows a maximum eccentricity of about $ 0.35$, whereas the maximum eccentricity varies between $ 0.32 $ and $ 0.4 $ during
one period of the oscillation.
In agreement with the previous results, the simulations with the small mass planets, 1 and 2\,$M_\mathrm{jup}$, do not show
any significant disk eccentricity. Beyond a planet mass of about 3-4\,$M_\mathrm{jup}$ the disk becomes eccentric. 
As expected, the more massive planets form a much wider gap in the disk,
and the eccentricity of the disk is much higher for the $ 4\,M_\mathrm{jup} $ and more massive planets.

\begin{figure}
	\centering
	\includegraphics[width=\columnwidth]{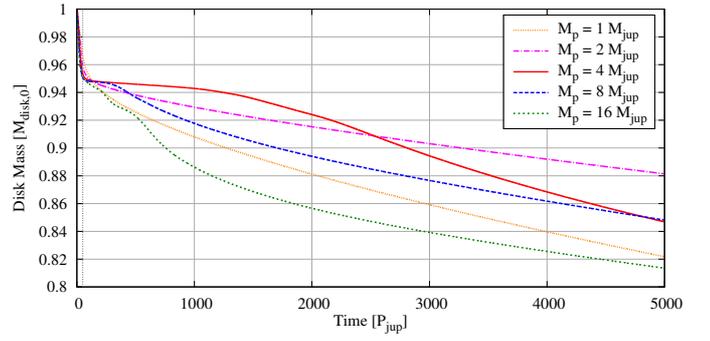} 
	\caption{
		\label{fig:mass_time_isothermal_mplanet}
		Total disk mass over time for the isothermal set-up with an inner radius of the computational domain of $ r_\mathrm{min} =0.7\,r_\mathrm{jup} $.
		The vertical dotted line at $ t = 50\,P_\mathrm{jup}$ marks the time when the planet has reached its full mass.
	}
\end{figure}

\begin{figure}
	\centering
	\includegraphics[width=\columnwidth]{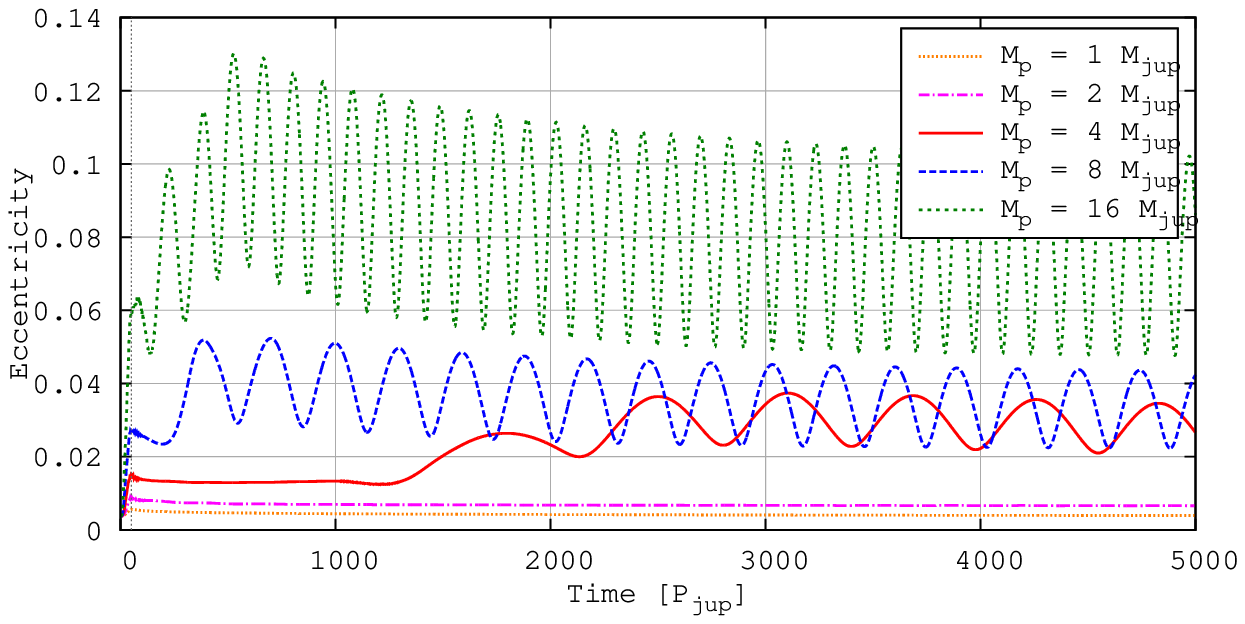} 
	\caption{
		\label{fig:eccentricity_time_isothermal_mplanet}
		Global disk eccentricity over time for the isothermal set-up with an inner radius of the computational domain of $ r_\mathrm{min} = 0.7\,r_\mathrm{jup} $.
		The vertical dotted line at $ t = 50\,P_\mathrm{jup}$ marks the time when the planet has reached its full mass.
	}
\end{figure}

\begin{figure}
	\centering
	\includegraphics[width=\columnwidth]{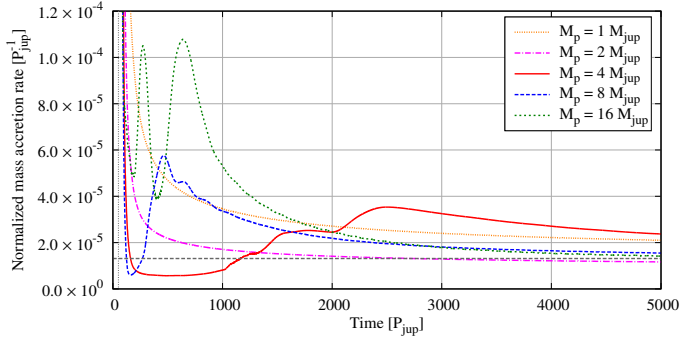} 
	\caption{
		\label{fig:massdot_time_isothermal_mplanet}
		Normalized mass accretion rate over time for the isothermal set-up with an inner radius of the computational domain of $ r_\mathrm{min} = 0.7\,r_\mathrm{jup} $.
		The values are plotted as a moving average over $ 50 $ orbits to remove jitter.
		The vertical dotted line at $ t = 50\,P_\mathrm{jup}$ marks the time when the planet has reached its full mass and the horizontal dashed line is the gas accretion rate given by Eq. (\ref{eq:accretionrate}).
	}
\end{figure}

In Figs.~\ref{fig:mass_time_isothermal_mplanet} and \ref{fig:eccentricity_time_isothermal_mplanet} the disk mass and 
disk eccentricity over time are shown. The average disk eccentricity increases clearly with the planetary mass.
It is easy to see that the higher the planet mass is, the faster the disk eccentricity starts to grow and that the mass loss starts earlier.
Whereas the previously examined example of the $ 4\,M_\mathrm{jup} $ planet needs about $ 1000 $ orbits until the disk eccentricity grows, the lower mass planet models do not develop a noteworthy eccentricity at all. 
The models with more massive planets start to develop much large eccentricities almost instantly. 
The normalized mass accretion rate is displayed in Fig.~\ref{fig:massdot_time_isothermal_mplanet}. 
There is no clear trend that with increasing planet mass the mass accretion rate grows or shrinks because there are two competing effects.
The growing disk eccentricity favours the mass transport across the planet into the inner cavity
of the disk whereas the larger gap due to the larger planet mass hinders mass transport through this gap. 
All the mass accretion rates are of the order of the typical disk accretion rates from Eq. (\ref{eq:accretionrate}).

\section{Radiative simulations}
\label{sec:radiative}

In Sect.~\ref{sec:isothermal} we performed locally isothermal simulations and
maintained the initial temperature stratification throughout the whole simulation. 
In this section we extend the set-up and also solve for the energy equation (Eq. \ref{eq:energy_equation}), but neglect
the radiative diffusion ($ 2 H \, \nabla \left( K \nabla T \right) = 0 $), because in standard accretion disk models
this term is much smaller than the vertical cooling. Additionally, we neglect stellar irradiation ($ Q_\mathrm{irrad} = 0 $).
We redid the same set of calculations (except for the $ r_\mathrm{min} = 0.4\,r_\mathrm{jup}$ case) as in the isothermal case for a direct comparison of the results.

\subsection{Inner radius dependency}

\begin{figure}
	\centering
	\includegraphics[width=\columnwidth]{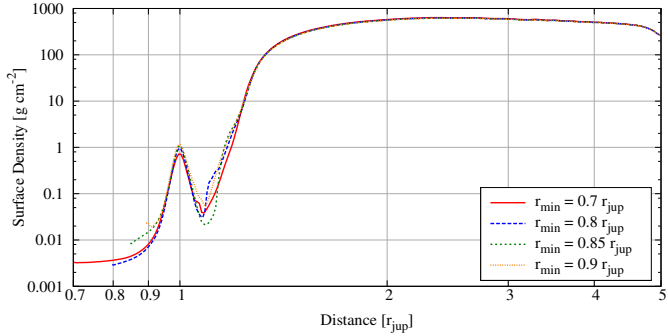} 
	\caption{
		\label{fig:sigma_radius_radiative_rmin_5000_log}
		Azimuthally averaged radial profile of surface density of the radiative set-up after $ 5000 $ orbits with a planet of $M_\mathrm{p} = 4\,M_\mathrm{jup} $.
	}
\end{figure}

\begin{figure}
	\centering
	\includegraphics[width=\columnwidth]{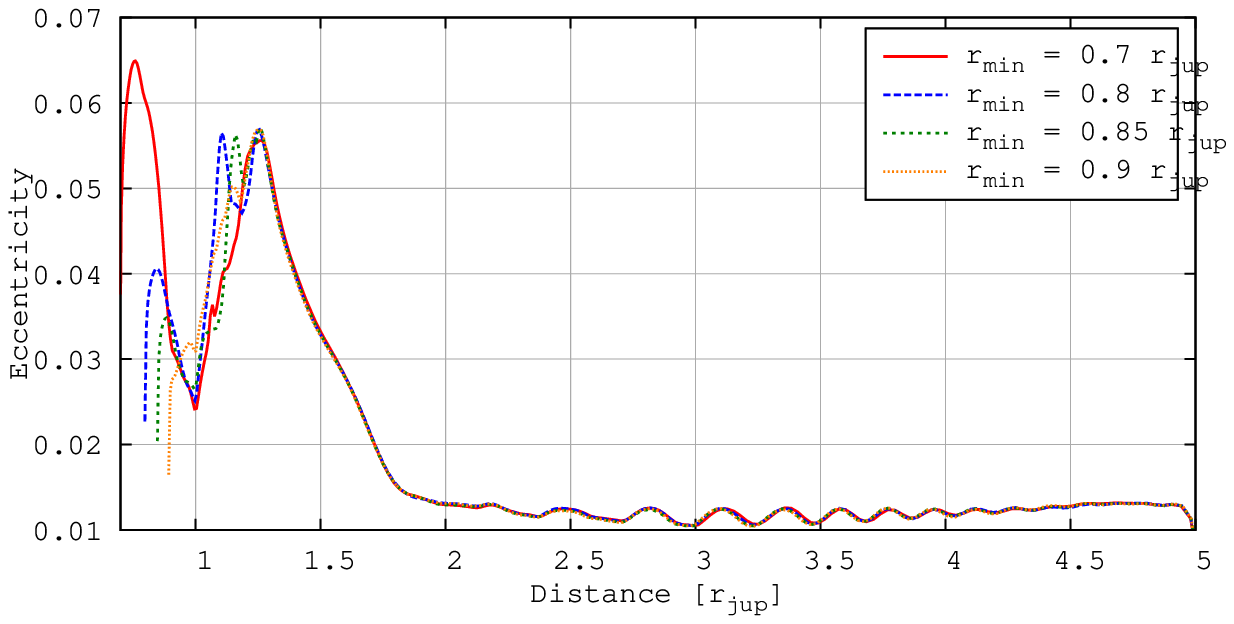} 
	\caption{
		\label{fig:eccentricity_radius_radiative_rmin_5000}
		Azimuthally averaged radial profile of eccentricity of the radiative set-up after $ 5000 $ orbits with a planet of $ M_\mathrm{p} = 4\,M_\mathrm{jup} $.
	}
\end{figure}

\begin{figure}
	\centering
	\includegraphics[width=\columnwidth]{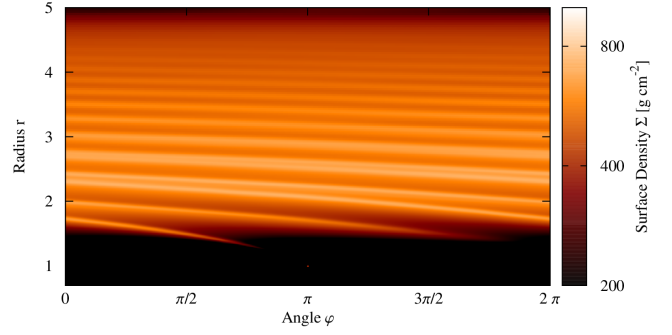} 
	\caption{
		\label{fig:sigma_radiative_5000}
		Surface density of the radiative set-up after $ 5000 $ orbits
		with a planet of $ M_\mathrm{p} = 4\,M_\mathrm{jup} $ and an inner radius of $ r_\mathrm{min} = 0.7\,\mathrm{jup} $. The planet is
		located at the angle of $ \pi $.
	}
\end{figure}

Figures~\ref{fig:sigma_radius_radiative_rmin_5000_log} and \ref{fig:eccentricity_radius_radiative_rmin_5000} show the surface density and eccentricity profiles at the end of the simulation.
Whereas the shape of surface density profile does not look too different from the isothermal models shown in Fig.~\ref{fig:sigma_radius_isothermal_rmin_5000_log}, 
the absolute values within the gap are about a magnitude smaller. All models display a much lower total eccentricity.
The planets also open a gap that can be seen in Fig.~\ref{fig:sigma_radiative_5000}, 
but compared with to the isothermal case (Fig.~\ref{fig:sigma_isothermal_5000}) the outer edge of the gap is much less
distorted because of the much smaller eccentricity. Only within the gap does a significant eccentricity develop and it is  a
factor of three to four smaller than in the isothermal simulations (see Fig.~\ref{fig:sigma_isothermal_5000}).

\begin{figure}
	\centering
	\includegraphics[width=\columnwidth]{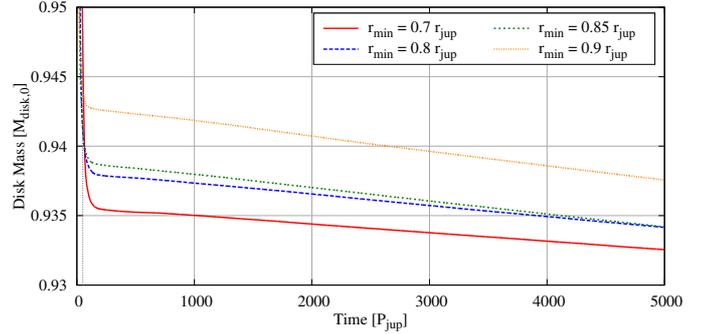} 
	\caption{
		\label{fig:mass_time_radiative_rmin}
		Total disk mass over time for the radiative set-up with a planet mass of $ M_\mathrm{p} = 4\,M_\mathrm{jup} $.
		The vertical dotted line at $ t = 50\,P_\mathrm{jup}$ marks the time when the planet has reached its full mass.
	}
\end{figure}

The global disk eccentricity is more or less constant over time 
and it reaches its final, very small value shortly after the planet has been ramped up. The value of $ 0.013 $ at the end of the simulation
is about the same as the value for the first stage in the isothermal evolution (Fig.~\ref{fig:eccentricity_time_isothermal_rmin}).
Because of the small constant value of the disk eccentricity there is a small mass loss through the inner edge of
the simulation region, which can be seen in the disk mass over time simulation shown in Fig.~\ref{fig:mass_time_radiative_rmin}.
As in the isothermal models, the mass loss at the start of simulation is dependent on the
exact position of the inner boundary, but for the rest of the simulation the results are only shifted.

\begin{figure}
	\centering
	\includegraphics[width=\columnwidth]{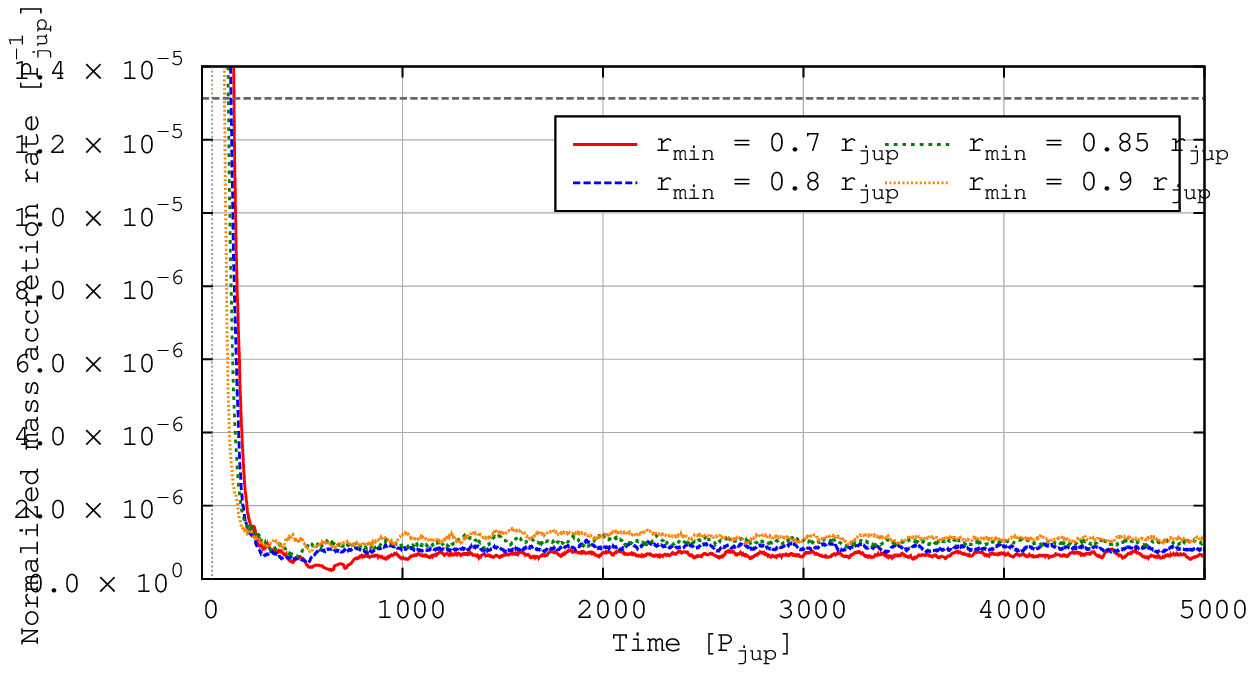} 
	\caption{
		\label{fig:massdot_time_radiative_rmin}
		Normalized mass accretion rate over time for the radiative set-up with a planet mass of $ M_\mathrm{p} = 4\,M_\mathrm{jup} $.
		The values are plotted as a moving average over $ 50 $ orbits to remove jitter.
		The vertical dotted line at $ t = 50\,P_\mathrm{jup}$ marks the time when the planet has reached its full mass and the horizontal dashed line is the gas accretion rate given by Eq. (\ref{eq:accretionrate}).
	}
\end{figure}

The normalized mass accretion rate shown in Fig.~\ref{fig:massdot_time_radiative_rmin} is therefore almost constant except for some jitter.
For the $ 0.7\,r_\mathrm{jup} $ model we obtain an average value of $ \left( 6.604 \pm 0.008 \right) \times 10^{-7} P_\mathrm{jup}^{-1} $ after $ 1000 $\,orbits.
Assuming a constant disk mass (see Eq. (\ref{eq:factor})), this corresponds to a mass accretion rate of $ \left( 6.680 \pm 0.008 \right) \times 10^{-9} M_{\sun}\,\mathrm{a}^{-1} $,
which fits very well to observational data \citep{2013ApJ...769..149K}.
Interestingly, it is now much smaller than the value given by Eq. (\ref{eq:accretionrate}) in contrast to the isothermal simulations,
as the disk is much less eccentric and therefore pushes less mass through the gap.
In addition, the disk can radiate its energy away and is therefore much cooler which also hinders mass transport through the gap.
In the next section we present a comparison of the temperatures in the isothermal and radiative disks.

\begin{figure}
	\centering
	\includegraphics[width=\columnwidth]{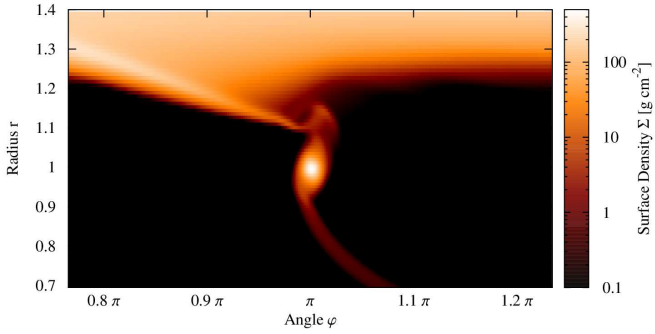} 
	\caption{
		\label{fig:sigma_radiative_5000_detail}
		Surface density of the radiative set-up after $ 5000 $ orbits
		with a planet of $ M_\mathrm{p} = 4\,M_\mathrm{jup} $ and an inner radius of $ r_\mathrm{min} = 0.7\,\mathrm{jup} $. 
               The planet is located at the angle of $ \pi $.
	}
\end{figure}

Figure~\ref{fig:sigma_radiative_5000_detail} shows a zoomed version of Fig.~\ref{fig:sigma_radiative_5000} with a different scaling of the surface density.
Within the gap almost no gas is left ($< 0.001\,\mathrm{g}\,\mathrm{cm}^{-2}$), but the spiral arm induced by the planet is clearly visible.

\subsection{Planet mass dependency}

\begin{figure}
	\centering
	\includegraphics[width=\columnwidth]{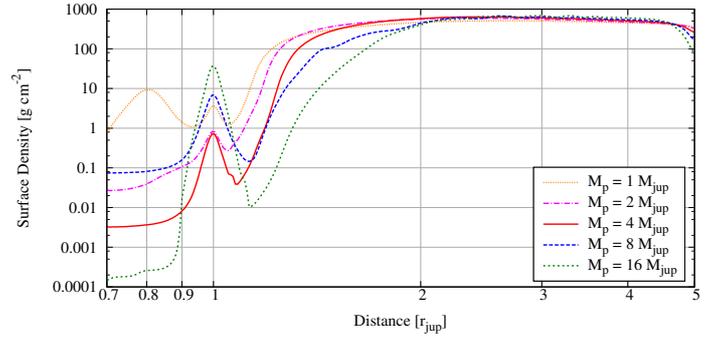} 
	\caption{
		\label{fig:sigma_radius_radiative_mplanet_5000_log}
		Azimuthally averaged radial profile of surface density of the radiative set-up after $ 5000 $ orbits with an inner radius of the computational domain of $ r_\mathrm{min} = 0.7\,r_\mathrm{jup} $.
	}
\end{figure}

\begin{figure}
	\centering
	\includegraphics[width=\columnwidth]{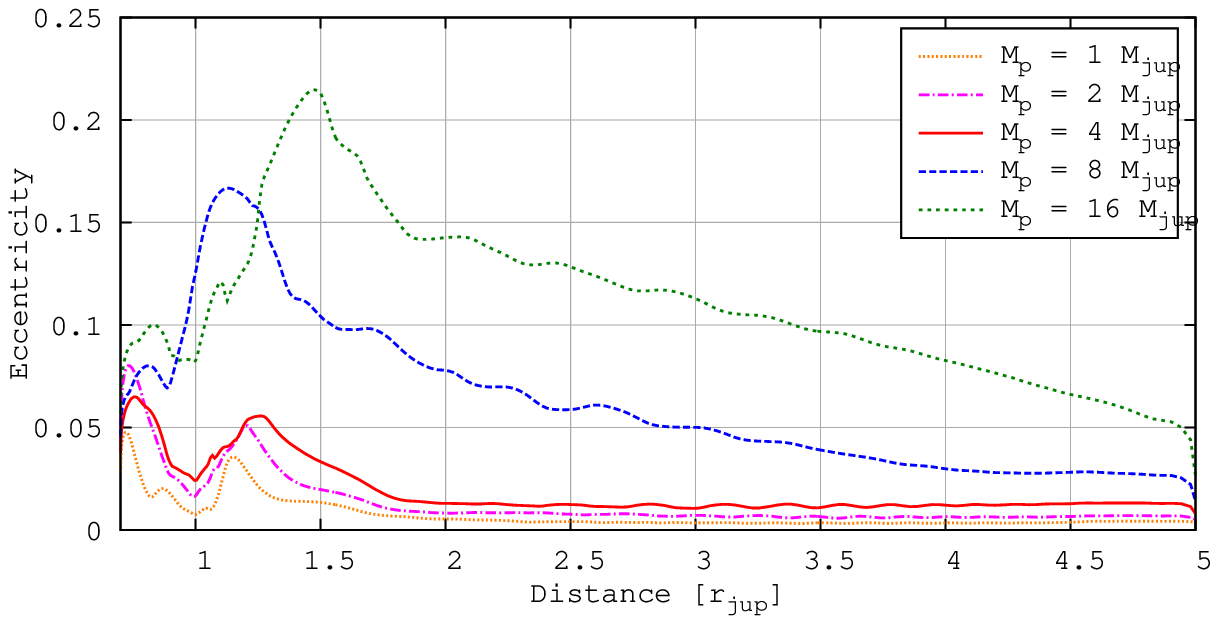} 
	\caption{
		\label{fig:eccentricity_radius_radiative_mplanet_5000}
		Azimuthally averaged radial profile of eccentricity of the radiative set-up after $ 5000 $ orbits with an inner radius of the computational domain of $ r_\mathrm{min} = 0.7\,r_\mathrm{jup} $.
	}
\end{figure}

\begin{figure}
	\centering
	\includegraphics[width=\columnwidth]{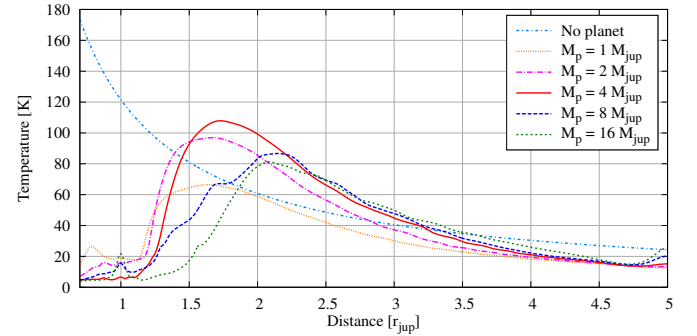} 
	\caption{
		\label{fig:temperature_radius_radiative_mplanet_5000}
		Azimuthally averaged radial profile of temperature of the radiative set-up after $ 5000 $ orbits with an inner radius of the computational domain of $ r_\mathrm{min} = 0.7\,r_\mathrm{jup} $.
	}
\end{figure}

For the comparison of different planet masses we again chose $ r_\mathrm{min} = 0.7\,r_\mathrm{jup}$ for the inner boundary. Figures~\ref{fig:sigma_radius_radiative_mplanet_5000_log} and
\ref{fig:eccentricity_radius_radiative_mplanet_5000} display the surface density and eccentricity profiles at the end of the simulation. As in the isothermal, case the more massive planets open a deeper gap in the disk.
In the isothermal simulations, the disk eccentricity profiles were much larger for masses $ M_\mathrm{p} \geq 4\,M_\mathrm{jup} $, whereas in the radiative models this is only the case for
masses $ M_\mathrm{p} \geq 8\,M_\mathrm{jup} $. In addition, all eccentricities are smaller than those seen in the isothermal case.
We note again, that the models are in different phases of their eccentricity oscillation as shown in Fig.~\ref{fig:eccentricity_time_radiative_mplanet}.
Because the temperature profile can also change during the simulation, Fig.~\ref{fig:temperature_radius_radiative_mplanet_5000} shows
the radial temperature profile at the end of the simulation.
The temperature in the gap is rather low and therefore we can see the deeper gaps for the more massive planets even in the temperature profile.
The maximum temperature is always at the outer edge of the gap
and is dependent on the planet's mass. The edge temperature is the highest for the $ 4\,M_\mathrm{jup} $
and lower for higher and smaller planet masses.

\begin{figure}
	\centering
	\includegraphics[width=\columnwidth]{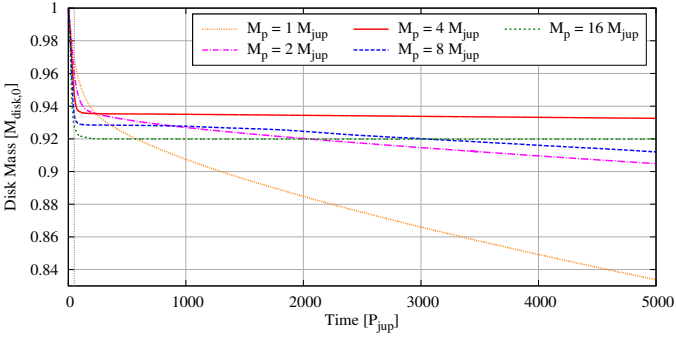} 
	\caption{
		\label{fig:mass_time_radiative_mplanet}
		Total disk mass over time for the radiative set-up with an inner radius of the computational domain of $ r_\mathrm{min} =0.7\,r_\mathrm{jup} $.
		The vertical dotted line at $ t = 50\,P_\mathrm{jup}$ marks the time when the planet has reached its full mass.
	}
\end{figure}

\begin{figure}
	\centering
	\includegraphics[width=\columnwidth]{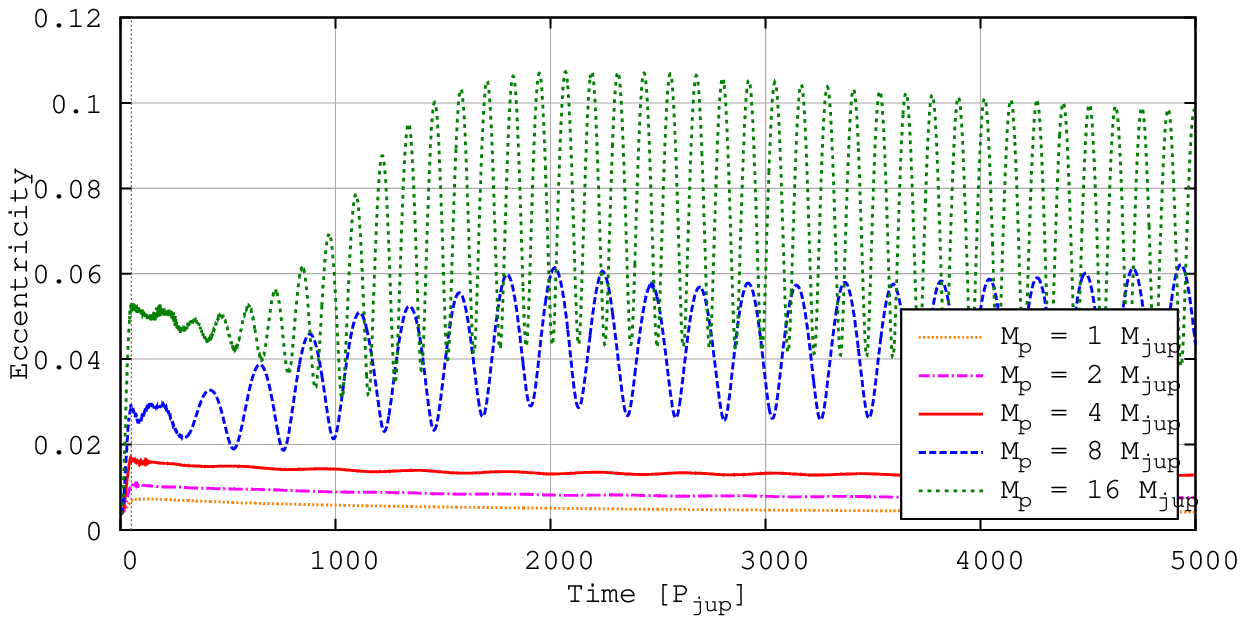} 
	\caption{
		\label{fig:eccentricity_time_radiative_mplanet}
		Global disk eccentricity over time for the radiative set-up with an inner radius of the computational domain of $ r_\mathrm{min} = 0.7\,r_\mathrm{jup} $.
		The vertical dotted line at $ t = 50\,P_\mathrm{jup}$ marks the time when the planet has reached its full mass.
	}
\end{figure}

\begin{figure}
	\centering
	\includegraphics[width=\columnwidth]{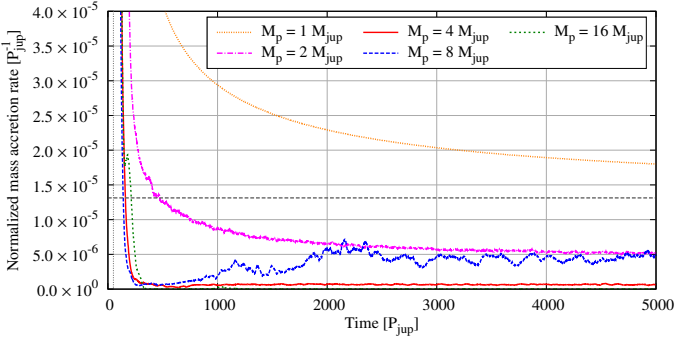} 
	\caption{
		\label{fig:massdot_time_radiative_mplanet}
		Normalized mass accretion rate over time for the radiative set-up with an inner radius of the computational domain of $ r_\mathrm{min} = 0.7\,r_\mathrm{jup} $.
		The values are plotted as a moving average over $ 50 $ orbits to remove jitter.
		The vertical dotted line at $ t = 50\,P_\mathrm{jup}$ marks the time when the planet has reached its full mass and the horizontal dashed line is the gas accretion rate given by Eq. (\ref{eq:accretionrate}).
	}
\end{figure}

The global disk mass evolution is shown in Fig.~\ref{fig:mass_time_radiative_mplanet}. 
The mass loss through the inner edge now depends on two factors. On the one hand, the mass flow through the gap
will be increased by the disk's eccentricity as in the isothermal case. The global disk eccentricity evolution in time as shown in Figure
\ref{fig:eccentricity_time_radiative_mplanet} is similar to that of the isothermal case. 
In the radiative case the planet mass required to reach a certain disk eccentricity is higher than in the isothermal case.
On the other hand thermal pressure changes the depth of the gap which may lead to an enhanced mass flow
across the gap.
This is also mass dependent as already seen in Fig.~\ref{fig:temperature_radius_radiative_mplanet_5000}.
Figure~\ref{fig:massdot_time_radiative_mplanet} now shows the normalized mass accretion rate for the radiative models. 
As both effects interact we can see no clear trend for the mass accretion rate. For the two lowest mass planets (1 and 2\,$M_\mathrm{jup}$),
the mass accretion rates are highest because they have the smallest gap size (see Fig.~\ref{fig:sigma_radius_radiative_mplanet_5000_log}).
The $4\,M_\mathrm{jup}$ case has a wider and deeper gap with no disk eccentricity and shows a much smaller accretion rate.
For the $8\,M_\mathrm{jup}$ case the disk becomes eccentric and the mass flow increases again, while for the $16\,M_\mathrm{jup}$ case
the gap widens substantially such that the mass flow rate becomes very small. 

\subsection{Disk mass dependency}

In the isothermal models the normalized mass accretion rate was independent of the disk mass $ M_\mathrm{disk} $ because as the surface density $ \Sigma $ cancels out of the equations. 
Therefore, the mass accretion rate scales with the same factor as the surface density, for example if we decrease the surface density by a factor of $ 10 $, the mass accretion rate decreases by a factor of $ 10$.

\begin{figure}
	\centering
	\includegraphics[width=\columnwidth]{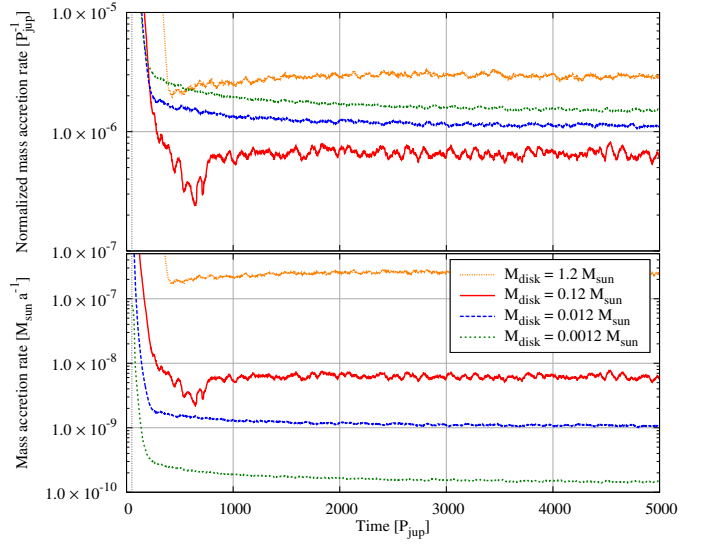} 
	\caption{
		\label{fig:massdot_time_radiative_mass_dependency}
		Normalized mass accretion rate over time (upper panel) and mass accretion rate over time (lower panel) 
		for the radiative set-up with an inner radius of the computational domain of $ r_\mathrm{min} = 0.7\,r_\mathrm{jup} $  and a planet mass of $ M_\mathrm{p} = 4\,M_\mathrm{jup} $.
		The values are plotted as a moving average over $ 50 $ orbits to remove jitter.
		The vertical dotted line at $ t = 50\,P_\mathrm{jup}$ marks the time when the planet has reached its full mass.
	}
\end{figure}

In the radiative case,  however, this is no longer true. Figure~\ref{fig:massdot_time_radiative_mass_dependency} shows the normalized mass accretion rates and the mass accretion rates for different disk masses.
The normalized mass accretion rates (upper panel of Fig.~\ref{fig:massdot_time_radiative_mass_dependency}) seem to scale in the wrong direction at first sight. For example, the $ M_\mathrm{disk} = 0.12\,M_{\sun} $ model
has a smaller normalized mass accretion rate than the $ M_\mathrm{disk} = 0.012\,M_{\sun} $ model, but if we calculate the mass accretion rates (lower panel of Fig.~\ref{fig:massdot_time_radiative_mass_dependency}) with Eq.~(\ref{eq:factor}) the trend
flips as expected and the  $ M_\mathrm{disk} = 0.12\,M_{\sun} $ model has a larger mass accretion rate than the $ M_\mathrm{disk} = 0.012\,M_{\sun} $ model, but it does not scale with the same factor as the disk mass.
For the rest of the simulations we used the $ M_\mathrm{disk} = 0.12\,M_{\sun} $ model.

\section{Comparison of different models}
\label{sec:comparison}

In this section we extend our models again for radiative diffusion and stellar irradiation and
compare the following four different models:
the isothermal set-ups from Sect.~\ref{sec:isothermal} with a constant temperature stratification throughout the simulation,
the radiative set-ups from Sect.~\ref{sec:radiative} where we solve the energy equation with viscous dissipation and radiative cooling,
the radiative with diffusion set-ups where we also solve the radiative diffusion in the $r$--$\varphi$ plane, and
the radiative with irradiation set-ups where we include radiative diffusion and stellar irradiation.
As radiative diffusion and stellar irradiation are very costly in terms of computation time we ran those models only with an
inner radius of $ r_\mathrm{min} = 0.7\,r_\mathrm{jup} $ and planet
masses of $ M_\mathrm{p}=4\,M_\mathrm{jup} $ and $ M_\mathrm{p}=8\,M_\mathrm{jup} $.

\begin{figure}
	\centering
	\includegraphics[width=\columnwidth]{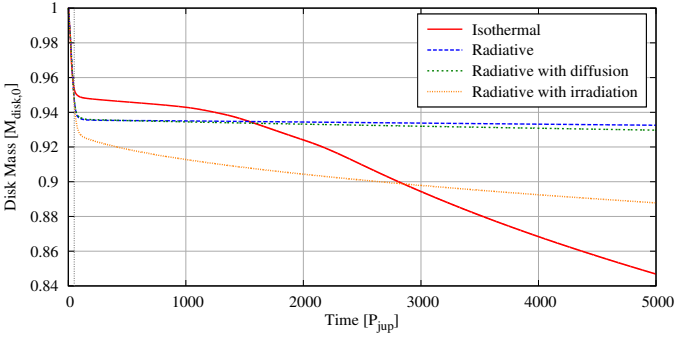} 
	\caption{
		\label{fig:mass_time_model_dependency_m4}
	Total disk mass over time for different set-ups with an inner radius of the computational domain
       of $ r_\mathrm{min} =0.7\,r_\mathrm{jup} $ and a planet mass of $ M_\mathrm{p} = 4\,M_\mathrm{jup} $.
		The vertical dotted line at $ t = 50\,P_\mathrm{jup}$ marks the time when the planet has reached its full mass.
	}
\end{figure}

\begin{figure}
	\centering
	\includegraphics[width=\columnwidth]{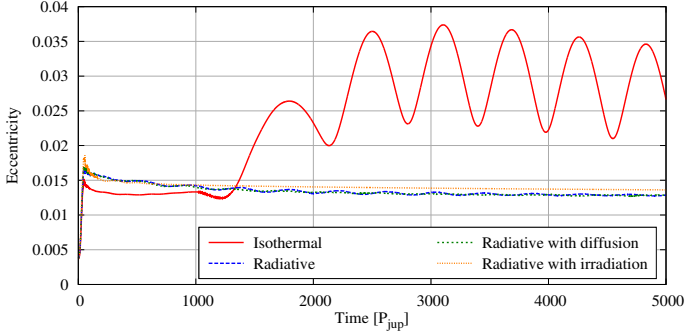} 
	\caption{
		\label{fig:eccentricity_time_model_dependency_m4}
		Global disk eccentricity over time for different set-ups with an inner radius of the computational domain of $ r_\mathrm{min} =0.7\,r_\mathrm{jup} $ and a planet mass of $ M_\mathrm{p} = 4\,M_\mathrm{jup} $.
		The vertical dotted line at $ t = 50\,P_\mathrm{jup}$ marks the time when the planet has reached its full mass.
	}
\end{figure}

Figures~\ref{fig:mass_time_model_dependency_m4} and \ref{fig:eccentricity_time_model_dependency_m4} show the disk mass and 
disk eccentricity evolution over the whole simulation for the $ M_\mathrm{p} = 4\,M_\mathrm{jup} $ models.
The isothermal model shows a huge difference in both plots. 
The disk loses much more mass and becomes significantly more eccentric. 
Between the radiative model and the radiative with diffusion model there is barely
a visible difference, so that the neglect of the radiative diffusion in accretion disk simulations is a valid assumption.
The addition of irradiation increases the mass loss again because the disk is much hotter,
but it does not play an important role for the global disk eccentricity. 

\begin{figure}
	\centering
	\includegraphics[width=\columnwidth]{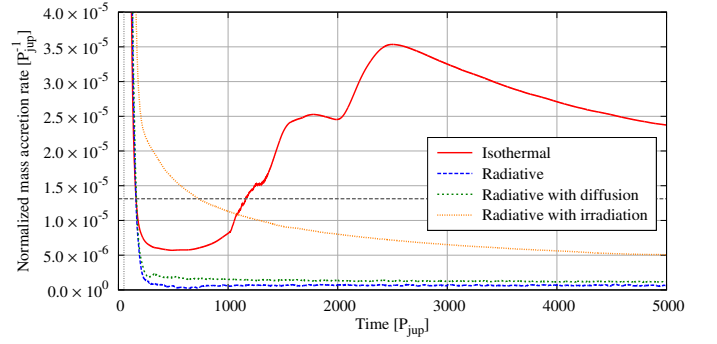} 
	\caption{
		\label{fig:massdot_time_model_dependency_m4}
		Normalized mass accretion rate over time for different set-ups with an inner radius of the computational
      domain of $ r_\mathrm{min} =0.7\,r_\mathrm{jup} $ and a planet mass of $ M_\mathrm{p} = 4\,M_\mathrm{jup} $.
		The vertical dotted line at $ t = 50\,P_\mathrm{jup}$ marks the time when the planet has reached its full mass and the horizontal dashed line is the gas accretion rate given by Eq. (\ref{eq:accretionrate}).
	}
\end{figure}

The normalized mass accretion rate of all models in shown in Fig.~\ref{fig:massdot_time_model_dependency_m4}. 
As already seen in Fig.~\ref{fig:mass_time_model_dependency_m4}, the mass loss of the isothermal model
is much larger than all the other models. 
The radiative with diffusion model has a slightly larger accretion rate than the radiative model, but both are smaller than the 
standard mass accretion rate of stationary accretion disks (Eq.~\ref{eq:accretionrate}). 
The accretion rate of the radiative with irradiation model needs much more time to converge
and is significantly larger than those of the other two radiative models.

\begin{figure}
	\centering
	\includegraphics[width=\columnwidth]{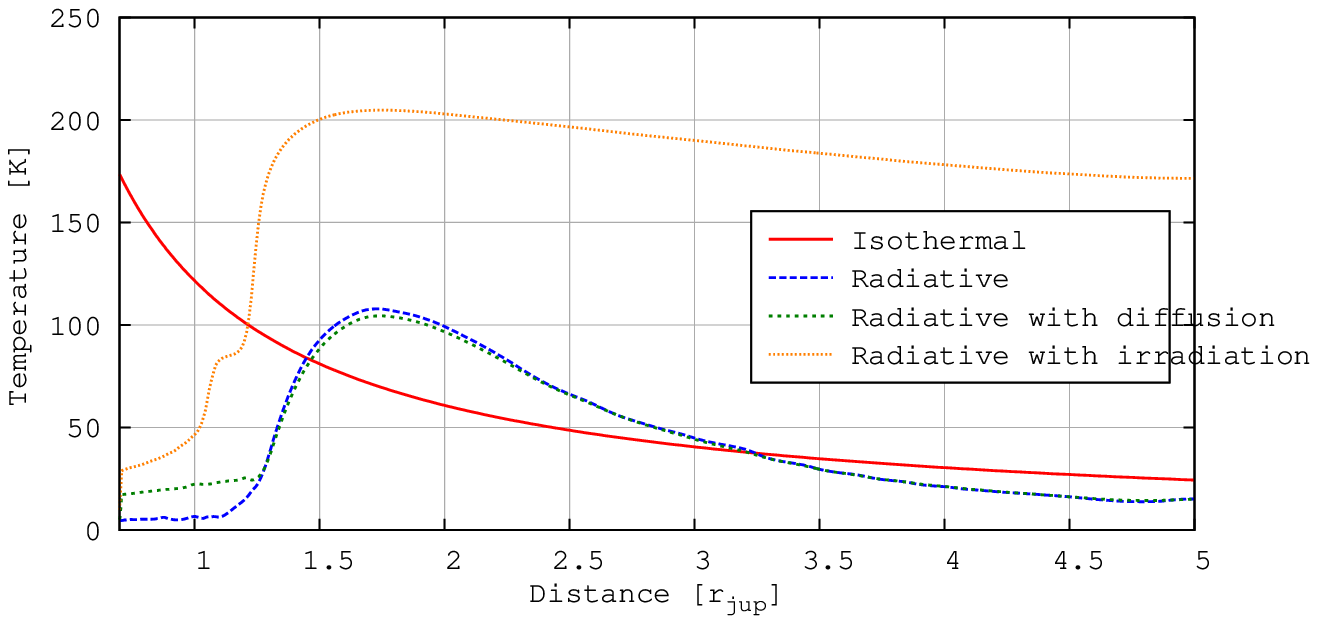} 
	\caption{
		\label{fig:temperature_radius_model_dependency_5000}
		Azimuthally averaged radial profile of temperature of the isothermal, radiative, radiative with diffusion and radiative with irradiation set-up
		after $ 5000 $ orbits with an inner radius of the computational domain of $ r_\mathrm{min} = 0.7\,r_\mathrm{jup} $ and a planet mass of $ M_\mathrm{p} = 4\,M_\mathrm{jup} $.
	}
\end{figure}

Therefore, the mass accretion in the irradiative model must be driven by thermal pressure through the gap, 
as the eccentricity is the same. This can be seen in Fig.~\ref{fig:temperature_radius_model_dependency_5000},
which shows the temperature profiles of all models after $ 5000 $ orbits.
The isothermal model still has the unchanged initial condition, whereas the radiative models show a temperature drop in the gap.
The radiative with diffusion set-up shows a slightly higher temperature in the gap as heat is diffused into the gap from the outer edge.
The temperature of radiative with irradiation is much higher overall, especially in the outer regions,
but the gradient on the outer edge of the gap is also much steeper and therefore has pushed more material through the gap.

\begin{figure}
	\centering
	\includegraphics[width=\columnwidth]{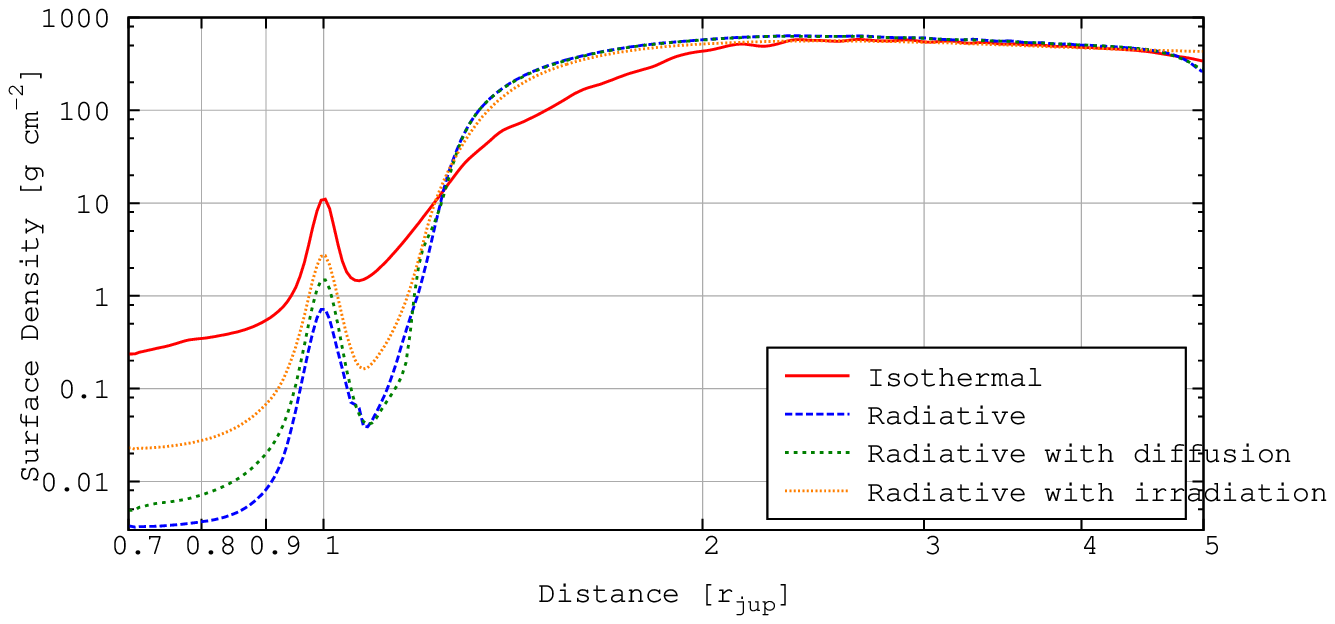} 
	\caption{
		\label{fig:sigma_radius_model_dependency_t=5000_log}
		Azimuthally averaged radial profile of the surface density of the 
      isothermal, radiative, radiative with diffusion and radiative with irradiation set-up
	after $ 5000 $ orbits with an inner radius of the computational domain of $ r_\mathrm{min} = 0.7\,r_\mathrm{jup} $ 
       and a planet mass of $ M_\mathrm{p} = 4\,M_\mathrm{jup} $.
	}
\end{figure}

In Fig.~\ref{fig:sigma_radius_model_dependency_t=5000_log}, we display the surface density at the final state
after 5000 orbits for all our models. In the outer part of the disk, all models show comparable surface densities.
In the inner gap region, the densities are typically 3 to 4 orders of magnitude smaller than in the outer disk.
As expected by the mass accretion rates, the isothermal and irradiated models show an enhanced density in the gap region.
In contrast, the radiative models have a surface density about an order of magnitude smaller inside the gap.
For the most realistic irradiated models, the density contrast is about 4 orders in magnitude.
For models with the 8\,$M_\mathrm{jup}$ planet the principal surface density structure is very similar for the radiative models.

\section{Summary and conclusions}
We have performed hydrodynamic simulations of transitional disks where the inner edge is formed by a very massive planet and studied the
mass flow rate of the gas from the outer disk into the inner cavity.
For comparison, we have studied locally isothermal disks as well as viscously heated radiative disks including irradiation from the
central star.

We find that the mass flow past the planet is largest for the isothermal models. These have a much higher temperature in the inner
regions of the disk and the corresponding pressure gradient is partly responsible for the high mass flow. The second contribution
comes from the eccentricity of the disk for planet masses beyond roughly $3\,M_\mathrm{jup}$.
The disk has its largest eccentricity near its inner edge, and reaches about $0.2$ for the $4$ and $8$\,$M_\mathrm{jup}$ planets and up to
0.35 for the $16\,M_\mathrm{jup}$ planet in the isothermal disk. For larger radii the disk eccentricity slowly declines.
For eccentric disks the planet passes periodically through the disk which leads to the
enhanced mass accretion into the inner gap.
The asymmetries in recent observations \citep[e.g.][]{2012ApJ...760L..26M} can be explained by these eccentricities or by a vortex within the disk
\citep{2013A&A...553L...3A}.

In the radiative cases, there are two reasons why the mass mass accretion rate is generally much smaller. 
First, the temperature is much smaller in the gap region which leads to a smaller pressure effect, and second the
radiative models have a smaller disk eccentricity than the isothermal models. The maximum disk eccentricity is around $0.2$ for
the most massive 16\,$M_\mathrm{jup}$ planet. 
The reduction in eccentricity comes about because the emission of radiation from the disk surfaces leads to an energy leakage
of the eccentric modes of the disk.
The model including stellar irradiation shows again a larger mass flow which is still smaller than the purely isothermal disk.

The surface density contrast between the outer and inner disk is interesting. In the eccentric disk case we find a contrast of 
about four orders of magnitude which is even larger in the non-eccentric disk cases.
Despite this large contrast in surface density, the mass accretion rate past the planet into the gap
changes by much less. For the isothermal models it is always comparable to the typical disk accretion rate 
(see Fig.~\ref{fig:massdot_time_isothermal_mplanet}) in contrast to the observations that show a reduction of about an order
of magnitude compared to stationary disks \citep{2013ApJ...769..149K}.  
However, for the more realistic radiative cases, the mass accretion rate can be reduced by over an order of magnitude,
if the disk is not eccentric, i.e. if the planet is smaller than about 5\,$M_\mathrm{jup}$. 
This mass flow rate is in good agreement with the observations \citep{2013ApJ...769..149K}.
For larger
mass planets and eccentric disks the mass flow rate is about 40\,\% of the standard equilibrium disk value
(see Fig.~\ref{fig:massdot_time_radiative_mplanet}).

Our model contains several restrictions that could be improved on in subsequent works.
The models are only 2D because 3D runs over this extended simulation time of several thousand orbits
are not feasible at the present time. We estimate that because of the large mass of the planets and the deep gaps that are
opened, the differences between 2D and 3D may be not too large. For Jupiter-mass planets the gaps are nearly indistinguishable
\citep{2001ApJ...547..457K} in 2D and 3D. Furthermore, the planets cannot accrete material from their surroundings. Because of
 the strong decline of the density in the gap, this effect may be more important for low mass planets as well. It may play a role,
however, in the eccentric disk case where it could lead to higher planet masses than otherwise reachable \citep{2006A&A...447..369K}. 

In our models we kept the planet fixed in a circular orbit, while it is known that the planets will tend to migrate
because of the torques of the outer disk.
For circular planets in non-eccentric disks, the migration will be directed inwards. However, the eccentric disk
for more massive planets will make the orbit the slightly eccentric and the migration rate slows considerably \citep{2006ApJ...652.1698D}.
For $e \geq 0.2$, the migration may be even outwards, but these high planetary eccentricities are only reached for very massive planets
beyond $10\,M_\mathrm{jup}$ \citep{2001A&A...366..263P}. However, all these results are only obtained using locally isothermal disks.
The question of planetary migration and possible eccentricity growth will have to be addressed in future simulations.
In our case, we might consider the simulated single planet being the outermost planet in a multiple system. 
Our results should then give a good approximation of the expected mass flow of the gas into the gap.

We also did not include any dust particles within the models.
\citet{2006MNRAS.373.1619R} showed that the outer edge of the gap acts as a filter for larger particles ($\gtrsim 10 $ \textmu m).
However, the simulations by \citeauthor{2006MNRAS.373.1619R} are only locally isothermal.
The question whether this filter mechanism also works within more realistic, radiative simulations will also have to be addressed in future simulations.

\begin{acknowledgements}
	Tobias M\"uller received financial support from the Carl-Zeiss-Stiftung. 
	Wilhelm Kley acknowledges the support of the German Research Foundation (DFG) through grant KL 650/8-2 within the Collaborative Research Group FOR 759: 
	The Formation of Planets: The Critical First Growth Phase.
	We thank Andres Carmona for very fruitful discussions. We thank the referee for providing constructive comments which helped to improve this paper.
	Most of the simulations were performed on the bwGRiD cluster in T\"ubingen, which is funded by the Ministry for Education and Research of Germany and
	the Ministry for Science, Research and Arts of the state Baden-W\"urttemberg, and the cluster of the
	Collaborative Research Group FOR 759.
\end{acknowledgements}

\bibliographystyle{aa}
\bibliography{references}

\end{document}